\documentclass[sigconf,nonacm]{acmart}

\AtBeginDocument{%
  }

\usepackage{listings}
\usepackage{xcolor}
\usepackage[most]{tcolorbox}
\usepackage{multirow}

\definecolor{codebg}{RGB}{247,247,247}
\definecolor{codeborder}{RGB}{220,220,220}
\definecolor{kw}{RGB}{0,0,160}
\definecolor{str}{RGB}{0,128,0}
\definecolor{com}{RGB}{120,120,120}

\lstdefinestyle{github}{
  backgroundcolor=\color{codebg},
  basicstyle=\ttfamily\small,
  frame=single,
  rulecolor=\color{codeborder},
  frameround=ffff,
  numbers=left,
  numberstyle=\tiny\color{com},
  stepnumber=1,
  numbersep=8pt,
  breaklines=true,
  tabsize=2,
  showstringspaces=false,
  keywordstyle=\color{kw}\bfseries,
  stringstyle=\color{str},
  commentstyle=\color{com}\itshape
}

\newtcolorbox{takeawaybox}{
  colback=yellow!10,
  colframe=yellow!40!black,
  coltitle=black,
  fonttitle=\bfseries,
  title=Takeaway,
  boxrule=0.8pt,
  arc=3pt,
  left=8pt,
  right=8pt,
  top=6pt,
  bottom=6pt
}

\usepackage{tikz}
\usepackage{textcomp}

\newcommand\copyrighttext{%
  \footnotesize \textcopyright 2026 IEEE. Personal use of this material is permitted.  Permission from IEEE must be obtained for all other uses, in any current or future media, including reprinting/republishing this material for advertising or promotional purposes, creating new collective works, for resale or redistribution to servers or lists, or reuse of any copyrighted component of this work in other works. 

  Accepted as paper at GreenArch Workshop, part of ICSA.}
\newcommand{\copyrightnotice}{%
\begin{tikzpicture}[remember picture,overlay]
\node[anchor=south,yshift=10pt] at (current page.south) {\fbox{\parbox{\dimexpr\textwidth-\fboxsep-\fboxrule\relax}{\copyrighttext}}};
\end{tikzpicture}%
}

\begin{document}

\title{A Curated List of Open-source Software‑only Energy Efficiency Measurement Tools: A GitHub Mining Study}

\author{Manuela Bechara Cannizza}
\affiliation{%
  \institution{Federal University of Technology - Paraná (UTFPR)}
  \city{Francisco Beltrão}
  \country{Brazil}
}
\email{manuelabechara@alunos.utfpr.edu.br}

\author{Michel Albonico}
\affiliation{%
  \institution{University of Southern Denmark (SDU)}
  \city{Vejle}
  \country{Denmark}}
\email{mical@mmmi.sdu.dk}

\begin{abstract}
Energy efficiency has become a growing concern in software development, leading to the need for tools designed to measure energy consumption.
While several energy measurement tools are available as open-source projects, their characteristics and adoption remain underexplored.
This work presents an empirical study based on a Mining Software Repositories (MSR) approach to identify, classify, and analyze software energy monitoring tools publicly available on GitHub.
We qualitatively analyzed an initial dataset of $585$ repositories to identify key design aspects, including measurement granularity and underlying design principles.
After this analysis, we retained $24$ repositories as relevant energy measuring software tools.
The qualitative analysis we conduct reveals a clear evolution from early CPU-centric and machine-level monitoring utilities toward more diverse tools that support multi-level granularity (process, container, and AI workload levels) and integrate emission estimation capabilities.
This study provides the first structured overview of open-source energy and emission measurement tools from an MSR perspective, which may be beneficial for software architects when designing energy-aware software.
\end{abstract}


\maketitle
\copyrightnotice

\section{Introduction}

\begin{figure*}
    \centering
    \includegraphics[width=0.6\linewidth]{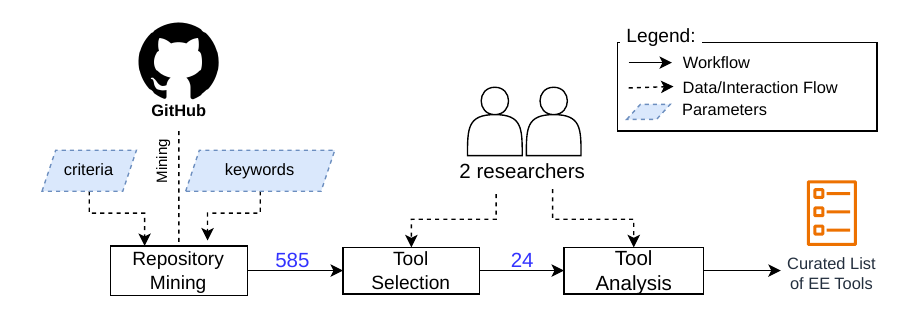}
    \caption{Workflow of the GitHub mining and data analysis.}
    \label{fig:workflow}
\end{figure*}

Energy efficiency has emerged as a critical non-functional requirement in modern software systems~\cite{ee-non-functional}. 
Software is present in different daily life activities, and the energy consumed by it plays a central role not only in extending the battery life of mobile devices but also in promoting sustainable digital practices.
This shift in focus has led to an increased demand for monitoring and evaluating the energy consumption of software in the development process and in-production usage~\cite{ee-software}.

There are different types of energy measurement approaches, the ones that rely on physical meter devices, and the software only ones that use already existing energy management~\cite{ee-software}, without the need of any physical device.
The software only energy measurement approaches align more with software development pipelines, with a facilitated integration as a linter, continuous integration (CI) and/or monitoring tool.
However, the landscape of these tools remains fragmented, and little is known about their design characteristics, maintenance status, or level of adoption within the developer community. 
A deeper understanding of these tools is essential to support both researchers and practitioners in selecting appropriate solutions and guiding the development of new energy-aware/efficient software architectures.

In this work, we present an empirical study following a Mining Software Repositories (MSR) approach to systematically identify, classify, and analyze software energy measurement tools publicly available on GitHub. 
We begin with a keyword-based search that retrieves $585$ repositories, from which we manually selected $24$ projects, based on their descriptions and filtering constraints, such as number of stars and watchers.
Each of the selected repositories undergoes a qualitative evaluation by two researchers to assess relevant design aspects, such as measurement granularity, sampling intervals, and the underlying measurement mechanisms. 
With this study, we aim to provide a curated list of the main software‑level tools for measuring energy consumption and related aspects at software development or deployment time. 

To achieve the research goal, we address the following research questions:

\noindent
\textbf{RQ1.} \emph{How has the development of energy and emission measurement tools evolved over time?}
By examining number of tools, we can identify the interest over time. 
Furthermore, with trends across years, we can contextualize current approaches and identify gaps that remain unaddressed.

\noindent
\textbf{RQ2.} \emph{What is the measurement granularity offered by existing tools?}
Granularity determines a tool’s suitability for specific tasks, such as profiling entire systems versus optimizing individual process/sub-process operations. Investigating granularity helps clarify which scenarios each tool is designed for and where further precision is needed.

\noindent
\textbf{RQ3.} \emph{What hardware and software restrictions or dependencies influence the adoption of current tools?}
Energy measurement may depend on specific hardware interfaces and software environments. 
Understanding these restrictions is crucial for assessing tool accessibility and portability, and for identifying barriers that limit widespread adoption.

Our findings provide a comprehensive overview of the current and most popular open-source energy monitoring tools.
The learned insights can inform future research and software development regarding energy-aware software engineering.
With this study, they will be able to make informed decisions on which type of tool they need to use, or even, which can be improved or further investigated.

\section{Software Energy Measurement}

Software energy measurement can be categorized along two main dimensions:
\emph{i)} hardware-assisted, where external devices, such as Power Distribution Units (PDUs)~\cite{pdu} and power meters~\cite{monsoon}, are used to measure computer machines energy consumption;
and \emph{ii)} software-only approaches, where the measurement relies on software layers that account for CPU, GPU, and DRAM usage mainly.
Hardware-assisted techniques are often considered the ground truth, as they capture the actual power draw at the device or system level with high accuracy and fine granularity, being suitable for controlled experiments and validation studies. However, they typically require specialized equipment, physical access to the infrastructure, and may not easily scale to large-scale or cloud-based environments.
In contrast, software-only approaches leverage built-in sensors and interfaces, such as Intel RAPL~\cite{rapl}, NVIDIA NVML~\cite{nvml}, and operating system counters, to estimate energy consumption with lower overhead and higher deployability. 

Software-only approaches enable continuous monitoring and integration into development workflows (e.g., CI/CD pipelines and IDEs), but may introduce estimation errors due to abstraction layers and limited visibility of certain hardware components~\cite{ee-measurement-accuracy}.
Additionally, software-based measurement can be further characterized by its granularity (e.g., system-level, process-level, or function-level profiling), which directly influence the trade-off with accuracy.

\section{Study Design}

Figure~\ref{fig:workflow} illustrates the workflow of our study, which is divided into 3 (three) main phases: \emph{(i)} \emph{repository mining}, \emph{(ii)} \emph{tool selection}, and \emph{(iii)} \emph{tool analysis}; 
which are detailed in sequence.

\begin{table*}[!htp]\centering
\caption{List of selected repositories as part of this study.}\label{tab:listrepos}
\begin{tabular}{lrrrrrrrr}\toprule
ID &repo\_name &year &stars &watchers &language &forks &open\_issues \\\midrule
\midrule
\multicolumn{8}{c}{Energy} \\
\midrule
R12 &hubblo-org/scaphandre &2020 &1900 &1900 &Rust &119 &103 \\
R14 &sustainable-computing-io/kepler &2022 &1458 &1458 &Go &223 &157 \\
R19 &trulyspinach/SMCAMDProcessor &2020 &1108 &1108 &C++ &102 &48 \\
R37 &jordond/jolt &2026 &333 &333 &Rust &11 &13 \\
R38 &ml-energy/zeus &2022 &332 &332 &Python &40 &15 \\
R50 &green-coding-solutions/green-metrics-tool &2022 &239 &239 &Python &44 &74 \\
R67 &graelo/pumas &2023 &198 &198 &Rust &7 &2 \\
R70 &ColinIanKing/powerstat &2015 &172 &172 &C &18 &1 \\
R86 &powerapi-ng/pyRAPL &2019 &117 &117 &Python &11 &10 \\
R88 &fvaleye/tracarbon &2022 &109 &109 &Python &7 &6 \\
R91 &joular/powerjoular &2021 &100 &100 &Ada &22 &10 \\
R94 &wattwisegames/watt-wiser &2024 &99 &99 &C++ &2 &0 \\
R96 &powerapi-ng/pyJoules &2019 &93 &93 &Python &15 &16 \\
R99 &joular/joularjx &2021 &92 &92 &Java &21 &3 \\
R440 &tdurieux/EnergiBridge &2023 &35 &3 &Python &27 &5 \\
R450 &enp1s0/gpu\_monitor &2020 &8 &8 &C++ &1 &0 \\
R475 &FairCompute/energy-monitoring-tool &2025 &7 &7 &Python &1 &5 \\
\midrule
\multicolumn{8}{c}{(Energy and) Emission} \\
\midrule
R13 &mlco2/codecarbon &2020 &1700 &1700 &Python &244 &128 \\
R32 &saintslab/carbontracker &2020 &473 &473 &Python &38 &14 \\
R46 &sb-ai-lab/Eco2AI &2022 &270 &270 &Python &23 &4 \\
R47 &mlco2/ecologits &2024 &255 &255 &Python &24 &11 \\
R163 &Root-Branch/cardamon-core &2023 &38 &38 &Rust &6 &30 \\
R483 &superdango/cloud-carbon-exporter &2025 &7 &7 &Go &1 &4 \\
R524 &NOVADEDOG/energy-leaderboard-runner &2025 &6 &6 &TypeScript &3 &2 \\
\bottomrule
\end{tabular}
\end{table*}

\subsection{Repository Mining}

In the first phase, we search for publicly available \emph{power} or \emph{energy} measurement and \emph{gas emission} estimation tools on GitHub.
This search has been orchestrated using the \emph{GitHub API\footnote{\url{https://docs.github.com/en/rest?apiVersion=2022-11-28}}}, with the following queries:

\begin{lstlisting}[language=bash, style=github, caption={GitHub API queries used to search for repositories.}, label={lst:gh-curl}]
    "energy measure is:public pushed:>2024-07-01 archived:false stars:>5 fork:false sort:stars",
    "energy monitor is:public pushed:>2024-07-01 archived:false stars:>5 fork:false sort:stars",
    "energy consumption is:public pushed:>2024-07-01 archived:false stars:>5 fork:false sort:stars",
    "power measur is:public pushed:>2024-07-01 archived:false stars:>5 fork:false sort:stars",
    "power monitor is:public pushed:>2024-07-01 archived:false stars:>5 fork:false sort:stars",
    "power consumption is:public pushed:>2024-07-01 archived:false stars:>5 fork:false sort:stars",
    "emission environment is:public pushed:>2024-07-01 archived:false stars:>5 fork:false sort:stars"
\end{lstlisting}

The queries consist of searches using three primary keywords, \emph{power}, \emph{energy}, and \emph{emission}, combined with secondary keywords (\emph{measure}, \emph{monitor}, \emph{consumption}, and \emph{environment}) to narrow the results to repositories relevant to the objectives of this research.
To ensure recent updates, we only consider repositories with a last push date after July~$1$,~$2024$ (6 months before this research study). 
The repository must also be public, guaranteeing access to the required information, and not archived, focusing on ongoing activity. 
Finally, we only select repositories with more than $5$ stars, looking for repositories that are minimally relevant to other projects.
Applying these criteria results in a total of $585$ repositories, which 
serve as the initial dataset for this study.
From each selected repository, we extract key attributes to guide the subsequent analysis: 
\emph{repository name}, \emph{description}, \emph{main topics}, \emph{creation date}, 
\emph{last update date}, \emph{size}, \emph{number of stars}, \emph{number of watchers}, 
\emph{primary programming language}, and \emph{number of forks}.

\subsection{Tool Selection}

During this phase, two researchers qualitatively analyzed the repositories, 
classifying them as either genuine power/energy measurement tools or not. 
The analysis involved visiting each repository and examining its description, \emph{README} file, 
and other relevant indicators. 
The two researchers independently reviewed all repositories, 
and the reliability of their classifications was assessed using Cohen's Kappa coefficient~\cite{cohen}, 
which yielded a value of $0.74$, indicating a \emph{substantial level of agreement}. 
Given that the list of repositories is not exhaustive, the researchers met to resolve 
discrepancies until reaching full consensus. 
As a result, only $29$ repositories were retained for further analysis. 
The excluded repositories were often toy projects, educational examples, hardware-dependent components (e.g., drivers), or unrelated to software energy measurement.

From the initial set of $29$ repositories, five appear at the bottom of the ranking table and have not been forked.
Upon closer inspection, these are personal projects with at most two collaborators, who seem to follow a development process without formal validation and with direct write access to the repositories.
Because these projects have attracted little attention until now, i.e., they have not been reused or incorporated into other software, we decided to exclude them from the final set of repositories analyzed in this study.
Table~\ref{tab:listrepos} presents the $24$ repositories\footnote{\url{https://docs.google.com/spreadsheets/d/16B3o5hvf0uXTuN8PBL5uNezSdOblUUAphoM2Jht4vDQ/edit?usp=sharing}} ultimately selected as software-level energy measurement tools.
In the table, the repositories are divided into two main classes: \emph{Energy} and \emph{(Energy and) Emission}. 
The \emph{Energy} group lists tools that exclusively focus on energy measurement, with creation years spanning from 2015 to 2026.
The \emph{(Energy and) Emission} group lists repositories that enable carbon and emission tracking.
Their creation years span between 2020 and 2025, and therefore, indicate a more recent interest in such a metric among software projects.

\subsection{Tool Analysis}

With the final dataset of $24$ power/energy measurement tool repositories, we conduct a second qualitative analysis when the two researchers meet together and classify together.
In this phase, we classify each project considering five main characteristics:
\emph{tool type}, \emph{granularity of the measurement}, \emph{frequency of the measurement}, \emph{software dependency}, and \emph{hardware dependency}.
These aspects determine both the scope and applicability of the tools. 
The \emph{tool type} indicates whether a repository provides a practical measurement tool, a dashboard, a development tool (such as to be used in the CI/CD pipeline), etc.
The \emph{granularity} of measurements (e.g., process, machine, or sub-process level) determines the level of detail the tool can measure. 
Similarly, the \emph{frequency} of data collection influences whether the tool captures fine-grained variations or is more suitable for long-term monitoring. 
The \emph{software dependency} reveals the potential integration with existing ecosystems. 
Finally, the \emph{hardware dependency} indicates the top-layer utilities dependency based on the machine architecture.

\section{Results and Discussion}

In this section, we present the main observations over the analyzed data, and further discuss their implications.
At the end of the section, we answer the research questions.

\subsection{Demographics and  Evolution}

\begin{figure}[!ht]
    \centering
    \includegraphics[width=\linewidth]{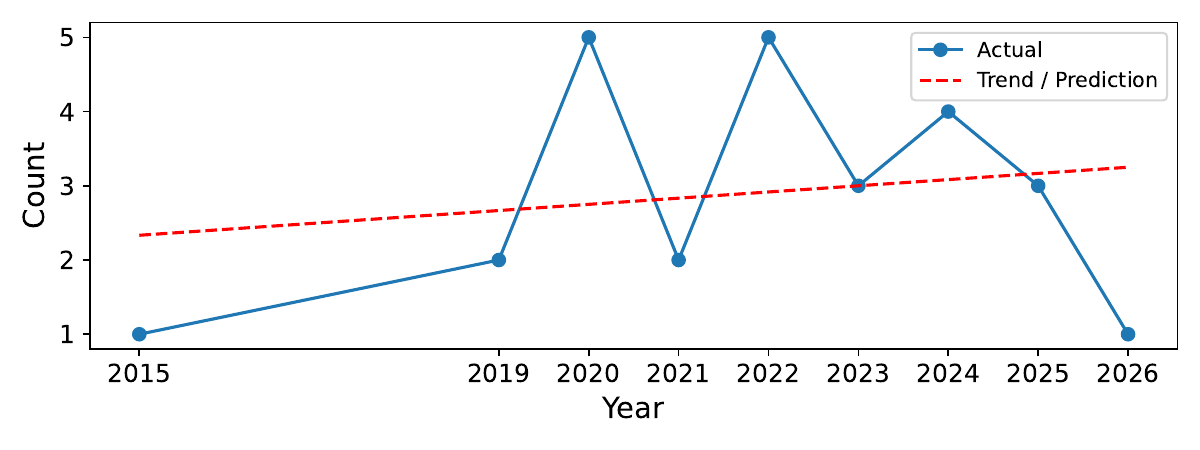}
    \caption{Number of measurement tools by year.}
    \label{fig:yearprediction}
\end{figure}

Figure~\ref{fig:yearprediction} shows the number of measurement tools created per year. 
Most tools have been developed since $2019$, indicating that there is a relatively recent movement within the software engineering community. 
The only repository created before $2019$ is \texttt{ColinIanKing/powerstat} (\emph{R70}), which dates back to $2015$, two years after Linux kernel support for Intel RAPL was first released.
Although $2026$ being considered in the regression calculation, considering it is the current and therefore incomplete year, the trend line suggests continued growth in the creation of such repositories. 
An important observation is the $2026$ project, which despite recent creation, is already among the most active in terms of stars, watchers, forks, and open issues, indicating its proeminence in the next years.
It is a terminal-based tool, similar to the Linux \texttt{top} utility, designed to identify processes with the highest energy consumption, which seems to be useful and should be closely observed in the near future.

To characterize the evolution of energy efficiency measurement tools, we conducted a \emph{thematic analysis}~\cite{thematic}, in the following sequence:
\emph{i)} both authors independently performed open coding on the repositories, proposing representative tokens that capture each tool’s primary technical focus (e.g., hardware scope, integration layer, deployment context, AI-awareness).
\emph{ii)} we compared and consolidated the initial codes through consensus meetings, resolving disagreements and refining token definitions.
\emph{iii)} tokens were grouped a posteriori into higher-level clusters via \emph{axial coding}~\cite{axial}, forming conceptually coherent categories.
\emph{iv)} we quantified the annual occurrence of tokens and analyzed their temporal distribution to identify dominant trends and transition periods.

Based on the convergence of thematic clustering and temporal patterns, we identify four evolutionary phases: \emph{hardware-centric}, \emph{developer libraries}, \emph{cloud-native}, and \emph{AI/LLM-aware}.
Table~\ref{tab:evolution-phases} lists these phases, their periods and their main characteristics.
The evolution of software-level energy measurement tools reveals a clear transition from hardware-bound, CPU-centric utilities toward multi-layer and AI-aware software. 
Early tools such as \texttt{R70} focused almost exclusively on machine-level CPU measurements, with limited or no GPU support. In contrast, more recent tools systematically incorporate GPU monitoring and support a broader range of hardware vendors (Intel, AMD, Apple, NVIDIA).
We also observe that earlier tools concentrated primarily on whole-machine measurements, whereas newer solutions enable more fine-grained instrumentation through language-specific libraries and wrappers (e.g., \texttt{R86} and \texttt{R47}).
From $2022$ onward, tools such as \texttt{R14} and \texttt{R50} reflect a DevOps and cloud-native perspective, incorporating container/pod-level granularity, dashboards, APIs, and Kubernetes integration. More recently ($2023$--$2025$), tools including \texttt{R47} and \texttt{R524} demonstrate a paradigm shift toward generative AI and LLM inference monitoring, introducing token- or prompt-level granularity and hybrid hardware-plus-model estimation approaches, often relying heavily on GPU measurements.
Finally, there is a conceptual shift from pure energy measurement to integrated energy-and-emissions accountability. Recent tools also estimate carbon footprint by incorporating CO$_2$ intensity models (e.g., \texttt{superdango/cloud-carbon-exporter}). Overall, this progression evidences the growing maturity of the community, both in technical sophistication and in its awareness of software’s contribution to greenhouse gas emissions.

\begin{table}[h!]
\centering
\caption{Phases of evolution of software-level energy measurement tools.}
\label{tab:evolution-phases}
\begin{tabular}{p{2cm}cp{4.3cm}}
\toprule
\textbf{Phase} & \textbf{Years} & \textbf{Characteristics} \\ 
\toprule
Hardware-centric & 2015--2019 & CPU-focused, machine-level measurements, terminal interface, limited GPU support. \\ \hline
Developer libraries & 2020--2021 & Wrappers for process/sub-process measurements, first GPU monitoring, energy-only or early emissions estimation. \\ \hline
Cloud-native & 2022--2023 & Container/pod-level granularity, dashboards and APIs, Kubernetes integration, multi-hardware support. \\ \hline
AI/LLM-aware & 2023--2025 & Token- or prompt-level measurements, hybrid CPU+GPU monitoring, CO$_2$ emission modeling, AI/LLM monitoring, richer interfaces. \\ 
\bottomrule
\end{tabular}
\end{table}

\subsection{Granularity of Measurements}

In this study, we consider granularity to the level at which energy or emission data is attributed (e.g., machine-level, process-level, container-level, prompt-level).
It fundamentally determines the \textit{actionability} and \textit{precision} of the collected energy or emission data.
Coarse-grained measurements (e.g., machine-level) provide a global view of energy consumption but make it difficult to identify which software component is responsible for inefficiencies. 
In contrast, fine-grained measurements (e.g., process-, container-, or prompt-level) enable developers to attribute energy usage to specific units of execution, allowing targeted optimizations such as refactoring, configuration tuning, or workload redistribution.

\noindent
\emph{Machine-level granularity.}
The majority of tools provide measurements at the \textbf{machine level}. 
Early utilities such as \emph{R70} and Apple-oriented monitors such as \emph{R67} focus exclusively on whole-machine energy consumption. 
More recent tools such as \emph{R94} and \emph{R37} maintain this approach, primarily targeting desktop or server monitoring scenarios. 
In total, machine-level granularity appears in more than half of the analyzed tools, indicating that hardware-level aggregation remains the dominant strategy.

\noindent
\emph{Process and sub-process granularity.}
A significant portion of tools provide \textbf{process-level} or \textbf{sub-process-level} measurement. 
Libraries such as \emph{R86}, \emph{R96}, and \emph{R91} allow developers to attribute energy consumption to specific code regions or program executions. 
Similarly, \emph{R12} and \emph{R50} offer per-process tracking, enabling more fine-grained attribution than whole-machine monitoring.
This level of granularity is particularly common in developer-oriented and research-oriented tools.

\noindent
\emph{Container, pod, and node granularity.}
With the rise of cloud-native systems, some tools extend granularity to \textbf{containerized environments}. 
For instance, \emph{R14} supports container-, pod-, and node-level energy metrics within Kubernetes clusters. 
This represents a shift from host-based measurement to infrastructure-aware attribution, aligning with modern DevOps and cloud deployment practices.

\noindent
\emph{Infrastructure-level granularity.}
A smaller subset of tools operates at the \textbf{infrastructure or cloud-resource level}. 
For example, \emph{R483} estimates energy and emissions for cloud services rather than individual processes or machines. 
This abstraction typically relies on energy and CO\textsubscript{2} models instead of direct hardware counters.

\noindent
\emph{LLM prompt and token-level granularity.}
The most recent tools introduce \textbf{AI-specific granularity}. 
\emph{R47} and \emph{R524} provide measurements at the level of LLM prompts or tokens, representing the finest abstraction observed in the dataset. 
This reflects the increasing importance of generative AI workloads and the need to attribute energy consumption at inference time.

\subsection{Software Dependencies}

Another important dimension in our analysis concerns the \textit{software dependency requirements}, which refers to the operating systems, runtime environments, middleware, or orchestration platforms required for a tool to function (e.g., Linux, macOS, Docker, Kubernetes, Python runtime).

\begin{table}
\centering
\caption{Project Dependencies on Software}
\label{tab:software-dependency-ids}
\begin{tabular}{p{3cm} p{4.8cm}}
\toprule
\textbf{Dependency Category} & \textbf{Project IDs} \\
\toprule
Linux OS &
R70, R86, R96, R12, R91, R99, R14, R38, R50, R314, R475, R37 \\

macOS &
R19, R67, R88, R440, R37 \\

Windows OS &
R14, R94, R440 \\

Cross-platform (2+ OS) &
R12, R13, R14, R88, R163, R440, R94, R37 \\

Python runtime &
R86, R96, R13, R32, R38, R46, R50, R88, R314, R47, R524 \\

Docker-based &
R12, R50, R14, R163, R483 \\

Kubernetes-dependent &
R14, R483 \\

Cloud / External API &
R13, R32, R46, R47, R483 \\

Machine-learning–specific &
R32, R38, R46, R47, R524 \\
\bottomrule
\end{tabular}
\end{table}

\noindent
\emph{Operating-System Dependency.}
A significant portion of tools depend directly on specific operating systems due to their reliance on hardware counters or system interfaces. 
Linux appears as the most common dependency, particularly for tools leveraging Intel RAPL or \texttt{nvidia-smi}. 
Several tools explicitly support multiple platforms (Linux, Windows, macOS), but cross-platform support often comes with limitations depending on hardware compatibility.
Tools targeting Apple Silicon (e.g., those relying on \texttt{powermetrics}) exhibit macOS-specific constraints, while some utilities depend on Windows-specific interfaces. 
Overall, OS dependency remains a strong limiting factor for reproducibility and portability.

\noindent
\emph{Programming Language and Runtime Dependency.}
Many tools are implemented as language-specific libraries, especially in Python. 
Examples include wrappers designed to instrument code execution (e.g., \emph{R86}, \emph{R88}, \emph{R96}), which require a language runtime and are therefore restricted to workloads implemented in that ecosystem.
This facilitates fine-grained instrumentation but reduces applicability to heterogeneous systems composed of multiple languages.

\noindent
\emph{Container and Orchestration Dependency.}
More recent tools introduce dependencies on containerization and orchestration platforms such as Docker (e.g., \emph{R50}) and Kubernetes (e.g., \emph{R13}). 
These tools operate at container- , pod- , or node-level granularity and integrate with cloud-native monitoring stacks (e.g., Prometheus exporters).
While this increases relevance for DevOps environments, it also introduces architectural lock-in: measurement becomes tightly coupled to specific deployment infrastructures.

\noindent
\emph{Cloud and API Dependency.}
Emission-oriented tools frequently depend on external carbon intensity models (e.g., \emph{R13}, \emph{R46}) or cloud APIs (e.g., \emph{R32}). 
Such dependencies enable location-aware or time-aware emission estimation but introduce external validity risks: changes in APIs, availability constraints, regional data inconsistencies, and
    model assumptions embedded in static CO\textsubscript{2} datasets.

\subsection{Hardware Specificities}

Hardware dependency refers to the extent to which energy and emission measurement tools rely on specific processor architectures, vendor interfaces, or hardware sensors (e.g., Intel RAPL, NVIDIA NVML/\texttt{nvidia-smi}, AMD \texttt{rocm-smi}, Apple \texttt{powermetrics}). 
 Our dataset indicates a transition from CPU-only, Intel-centric measurement (2015--2019) to heterogeneous hardware support (2020--2026), including GPUs and ARM-based processors. 
Nevertheless, hardware coupling remains a fundamental limitation, as most tools still rely on vendor-specific metering utilities rather than fully hardware-agnostic approaches.

Table~\ref{tab:cpuarch} lists the projects dependencies and the meter utility of each processing unit.

\begin{table}[!htp]\centering
\caption{Project Dependencies on Hardware}
\label{tab:cpuarch}
\begin{tabular}{p{1.5cm} p{1cm} p{2.0cm} p{2.5cm}}
\toprule
\textbf{Manu-facturer} & \textbf{Proc. Unit}  & \textbf{Meter Utility} & \textbf{Tools}  \\
\midrule
Intel & CPU & RAPL & 
R70, R86, R96, R12, R13, R32, R91, R99, R14, R38, R46, R50, R88, R440, R314, R94, R475, R524 \\
\midrule
AMD & CPU & RAPL & 
R70, R86, R96, R12, R13, R91, R99, R14, R38, R46, R50, R88, R440, R314, R94, R475, R524 \\
    &     & SMCAMD-Processor & 
R19 \\
    & GPU & rocm-smi & 
R13, R14, R88, R524 \\
\midrule
Apple & CPU & powermetrics\footnote{\url{https://ss64.com/mac/powermetrics.html}} & 
R88, R67, R524, R37 \\
      & GPU & powermetrics & 
R67, R524 \\
\midrule
NVIDIA & GPU & nvidia-smi\footnote{\url{https://developer.nvidia.com/system-management-interface}} & 
R96, R13, R32, R450, R91, R14, R38, R46, R88, R440, R314, R94, R475, R524 \\
\bottomrule
\end{tabular}
\end{table}

\noindent
\emph{CPU-Centric Dependency.}
The majority of tools depend primarily on CPU-based energy counters. 
Tools such as R70, R86, R96, R12, R13, R91, R99, and R50 leverage RAPL as their main measurement backend, indicating a strong reliance on x86 architectures. 
This dominance can be explained by the maturity, accessibility, and relatively low overhead of RAPL compared to alternative hardware counters. 
However, this also constrains applicability to platforms where RAPL is available.
Here, it is important to highlight that for all the projects relying on RAPL, we consider as supporting AMD processors since their most recent architectures have support to RAPL counters. 

\noindent
\emph{GPU Vendor Lock-in.}
GPU energy measurement shows a clear dependency on vendor-specific utilities. 
NVIDIA GPUs are predominantly monitored via \texttt{nvidia-smi}, which is used by several tools in the dataset (e.g., R96, R13, R32, R14, R38, R46, R94, R524). 
In contrast, AMD GPU support through \texttt{rocm-smi} appears in a much smaller subset of tools, highlighting an imbalance in ecosystem support. 
This asymmetry introduces a bias toward NVIDIA-based experimental setups in empirical energy studies.
Support for Apple hardware is primarily enabled through \texttt{powermetrics}, as observed in tools such as R67, R88, R37, and R524. 
While this expands coverage to ARM-based systems, it also introduces platform-specific constraints, since \texttt{powermetrics} is limited to macOS environments. 

\noindent
\emph{Multi-Hardware Support and Abstraction Layers.}
More recent tools attempt to mitigate hardware dependency by supporting multiple backends simultaneously (e.g., RAPL, \texttt{nvidia-smi}, \texttt{rocm-smi}, and energy models). 
Examples include R13, R88, R440, and R524, which integrate heterogeneous hardware interfaces to increase coverage across CPUs and GPUs. 
This trend suggests a shift toward abstraction layers that decouple measurement logic from specific hardware counters.

\subsection{Answering the Research Questions}

\noindent
\textbf{ARQ1.} \emph{How has the development of energy and emission measurement tools evolved over time?}
The development of energy and emission measurement tools shows a clear transition from hardware-centric and CPU-focused utilities to more heterogeneous and ecosys-tem-aware solutions. 
Early tools (2015--2019) primarily relied on low-level interfaces such as RAPL and provided machine-level energy measurements with limited scope and hardware coverage. 
From 2020 onward, tools increasingly incorporated GPU monitoring, emission estimation models, and dashboard-based visualization, reflecting the growing importance of AI and cloud workloads. 
The most recent tools (2023--2026) further extend this trajectory by supporting multi-platform environments, generative AI workloads, and integrated carbon estimation, indicating a shift from pure energy metering toward sustainability-oriented analytics and decision support.

\noindent
\textbf{ARQ2.} \emph{What is the measurement granularity offered by existing tools?}
The results reveal that machine-level granularity remains the dominant approach, particularly in system monitoring and hardware-oriented tools.
Nevertheless, there is a progressive diversification toward finer-grained measurement levels. 
Several tools now support process- and sub-process-level attribution, enabling developers to associate energy consumption with specific executions or code segments. 
More recent tools introduce container-, pod-, and infrastructure-level granularity aligned with cloud-native architectures, while a small but emerging subset targets AI-specific units such as model execution, prompts, or tokens. 
This evolution demonstrates a shift from coarse descriptive monitoring to more actionable and context-aware measurement that better supports optimization and empirical evaluation.

\noindent
\textbf{ARQ3.} \emph{What hardware and software restrictions or dependencies influence the adoption of current tools?}
The analysis highlights strong hardware and software dependencies that significantly influence tool adoption and applicability. 
On the hardware side, most tools rely on vendor-specific metering interfaces (e.g., RAPL for Intel/AMD CPUs, \texttt{nvidia-smi} for NVIDIA GPUs, \texttt{rocm-smi} for AMD GPUs, and \texttt{powermetrics} for Apple devices), creating platform-specific constraints and limiting cross-hardware comparability. 
On the software side, many tools depend on particular operating systems (predominantly Linux), language runtimes (especially Python), or execution environments such as Docker and Kubernetes. A
dditionally, emission-oriented tools often require external carbon intensity models or APIs. These dependencies introduce challenges for portability, reproducibility, and long-term maintainability, as measurements may vary depending on the underlying hardware architecture and software ecosystem in which the tools are deployed.

\section{Related Work}

A systematic search using combinations of the keywords \texttt{software}, \texttt{energy}, and \texttt{mining} did not reveal any prior study that comprehensively maps publicly available energy efficiency measurement tools hosted on GitHub. The closest related work is by Georgiou et al.~\cite{ee-techniques-tools}, who survey techniques and tools for measuring software energy efficiency; however, their analysis is restricted to peer-reviewed literature and does not account for gray literature or open-source repositories. Similarly, Mancebo et al.~\cite{ee-software} propose a systematic process for measuring the energy efficiency of software pipelines, yet they do not provide an in-depth investigation of the concrete tools that can support such a process. Moreover, several studies examine energy efficiency in specific domains, such as Green AI, ROS-based systems, Android applications, and data centers~\cite{slr-green-ai,ee-ros,ee-android, ee-datacenter}. While these works offer valuable insights into tool usage within their respective contexts, they do not present a consolidated overview of the available measurement tools. Therefore, our study addresses this gap by systematically identifying and mapping existing energy efficiency measurement tools in open-source ecosystems, providing a reference for future empirical analyses and informed tool selection.

\section{Threats to Validity}

In this section, we discuss threats to validity and describe the mitigation strategies adopted.

\subsection{Construct Validity}

The main construct validity threat lies in the definition of what qualifies as an \emph{energy measurement tool} since some repositories focus on dashboards, estimators, or auxiliary monitoring components rather than direct measurement. 
To mitigate this, we established explicit inclusion and exclusion criteria and performed a qualitative inspection of each repository, including its README, description, and documented functionality, and only kept the ones that provide some utility for energy measurement. 
Furthermore, two researchers independently reviewed the repositories and resolved disagreements through consensus meetings, increasing the robustness of the classification process. 

\subsection{Internal Validity}
One internal validity threat is the keyword-based repository mining strategy, which may omit relevant tools that do not explicitly use the selected terms. 
To reduce this risk, we employed multiple complementary query formulations and combined primary and secondary keywords when using the GitHub API, thereby broadening the search space and reducing query bias.

Another internal validity threat arises from manual qualitative coding and thematic analysis of repositories, which can be influenced by researcher subjectivity. 
We mitigated this threat through independent coding, iterative discussions, and consensus-based axial coding, following a structured thematic analysis procedure.

\subsection{External Validity}

Our study focuses exclusively on open-source repositories hosted on GitHub, which may not represent proprietary, industrial, or non-GitHub tools. 
Consequently, the curated list should be interpreted as representative of the open-source, as suggested in the paper ecosystem, as suggested in the paper title, rather than the entire landscape of energy measurement solutions.

\subsection{Conclusion Validity}

Since our results are primarily descriptive and exploratory, a threat lies in overinterpreting trends such as the evolution toward AI-aware or cloud-native tools. To mitigate this, we conduct multiple classifications, including temporal frequency analysis, thematic clustering, and dependency analysis across hardware and software dimensions.

Additionally, the relatively small final dataset of curated tools, after rigorous filtering, may limit statistical generalization. 
However, this is an inherent trade-off of in-depth qualitative MSR studies. 
We addressed this limitation by providing transparent selection criteria, documenting the full workflow, and grounding interpretations in observable repository characteristics.

\section{Conclusion and Future Work}

This paper presented an MSR-based empirical study of open-source energy and emission measurement tools available on GitHub, resulting in a curated dataset of 24 relevant repositories. 
Our analysis showed that the ecosystem has evolved from early, CPU-centric and machine-level monitoring utilities to more sophisticated tools that support finer-grained measurements, heterogeneous hardware, and sustainability-oriented features such as emission estimation and dashboard integration. 
Despite these advances, current tools remain strongly constrained by hardware and software dependencies, particularly vendor-specific interfaces (e.g., RAPL and \texttt{nvidia-smi}) and platform-specific environments, which limit portability and reproducibility across systems. 
Furthermore, while granularity has improved with the emergence of process-, container-, and AI-level measurement, machine-level monitoring still dominates, indicating a gap between available tooling and the needs of modern cloud-native and AI-driven software systems. 
Our findings highlight the need for more hardware-agnostic, standardized, and fine-grained measurement solutions to better support Green Software Engineering practices and enable reliable, comparable energy assessments across diverse computing environments.

As future work, we plan to empirically study the issues of those repositories, highlighting the main challenges and impediments in improving those solutions.
With this, we aim at making software developers and researchers aware of the next steps they can take in helping to evolve this important field in software engineering.

\section*{Acknowledge}

AI-based tools, particularly Generative AI systems, were used solely to support auxiliary tasks, including grammar correction, generation of \LaTeX{} code for tables based on our dataset, and drafting straightforward descriptive summaries of those tables. 

This research has been conducted while Michel Albonico held an Associated Professor position at the UTFPR, and therefore, that University should also get the credits of this publication.

Manuela Bechara Cannizza conducted the main part of the experiments as an intern of the PIBIC program.

\bibliographystyle{ACM-Reference-Format}

\bibliography{references}

\end{document}